\documentclass{iaus}
  
\usepackage{graphicx}

\title[Dwarf Ellipticals in Clusters] %% give here short title %%
{Constraining Galaxy Formation Models with Dwarf Ellipticals in Clusters}

\author[Christopher J. Conselice]   %% give here short author list %%
{Christopher J. Conselice$^1$}

\affiliation{$^1$ California Institute of Technology, Pasadena CA, USA}

\pubyear{2004}
\volume{xxx}  %% insert here IAU Symposium No.
\pagerange{119--126}
\date{?? and in revised form ??}
\jname{Proceedings Title IAU Symposium}
\editors{A.C. Editor, B.D. Editor \& C.E. Editor, eds.}
\begin{document}

\maketitle

\begin{abstract}

Recent observations demonstrate that dwarf elliptical (dE) galaxies in 
clusters, despite their faintness, are likely a critical galaxy type for 
understanding the processes behind galaxy formation. Dwarf 
ellipticals are the most common galaxy type, and are particularly abundant 
in rich galaxy clusters.  The dwarf to giant ratio is in fact highest in 
rich clusters of galaxies, suggesting that cluster dEs do not form in 
groups that later merge to form clusters. Dwarf ellipticals are potentially 
the only galaxy type whose formation is sensitive to global, rather than 
local, environment.  The dominant idea for explaining the formation of these 
systems, through Cold Dark Matter models, is that dEs form early and within 
their present environments.   Recent results suggest that some dwarfs appear 
in clusters after the bulk of massive galaxies form, a scenario not 
predicted in standard hierarchical structure formation models. Many dEs 
have younger and more metal rich stellar populations than dwarfs in lower 
density environments, suggesting processes induced by rich clusters play an 
important role in dE formation.  Several general galaxy cluster observations, 
including steep luminosity functions, and the origin of intracluster light, 
are natural outcomes of this delayed formation.

\end{abstract}

\firstsection % if your document starts with a section,
\section{Introduction}

Although dwarf galaxies are the faintest and lowest mass galaxies in the 
universe, they likely hold important clues for 
understanding galaxy formation and the nature of dark matter.  
The reason for this is quite simple: low-mass galaxies, and particularly 
low-mass galaxies in clusters,   are the most common 
galaxies in the nearby universe (Ferguson \& Binggeli 1994).  
Any ultimate galaxy evolution/formation theory must be able
to predict and accurately describe the properties of these objects.  
In popular galaxy formation models, such as hierarchical assembly (e.g., 
Cole et al. 2000), massive 
dark halos form by the mergers of lower mass ones early in the universe, and
the first galaxies likely have low stellar mass.  By
understanding dwarf galaxies, we are also 
thus potentially studying the very first
galaxies to form.   On the other hand, observations
reveal that few low-mass galaxies could have formed all of their stars 
early in the universe at $z > 7$,
with considerable evidence for star formation occurring in the last
few Gyrs in dwarf spheroidals (e.g., Mateo 1998).

The traditional approach to studying low-mass galaxies is to examine 
those in the Local Group (LG). It is now well established that LG dwarf
elliptical and dwarf spheroidal galaxies have varying star formation 
histories, with metal-poor populations as old as classical halo globular 
galaxies but also with evidence for recent
star formation (see e.g., Mateo 1998). There are also some low-mass LG 
galaxies, such as Sagittarius, that contain surprisingly metal-rich 
populations given their luminosities (e.g., Ibata, Gilmore \& Irwin 1995). 
Because LG dE galaxies are close we can resolve their stellar populations, 
and thus we can learn much more about them than we do dwarfs in more
dense, but distant environments.  Because of this we know a great
deal concerning LG dwarf properties including their 
internal kinematics and star formation histories. Many of the lowest mass 
galaxies in the LG, such as Draco and Ursa Minor, have very high inferred 
central M$_{\rm tot}$/L ratios (Kleyna et al. 2002) and apparently contain the 
densest dark matter halos of all known galaxies, in qualitative agreement 
with the original CDM predictions (e.g., Lake 1990). Low-mass galaxies in 
clusters, however, have different kinematic and spatial properties than 
these LG systems (e.g.,  Conselice, Gallagher, \& Wyse 2001,2003), 
suggesting they might have a different formation scenario.

Observationally, dwarf galaxies are typically faint (M$_{\rm B} > -18$), with
low
surface brightnesses ($\mu_{\rm B} > 23$ mag arcsec$^{2}$).  Since dwarfs are
so common, they are in every sense {\rm normal} galaxies.  The most
common type among dwarfs are dwarf ellipticals/spheroids, which
dominate the number density of galaxy clusters down to M$_{\rm B} = -11$
(Ferguson \& Bingelli 1994; Trentham et al. 2002).  Based on studies
of luminosity functions in clusters, there are also more dwarf ellipticals 
per giant in denser regions than in the field.  This implies that clusters of
galaxies cannot form through simple mergers of galaxy groups.  
Some dwarf systems must form within the cluster
environment. The nature of this over-density may be the result of initial
conditions, or `non-standard' galaxy formation. That is, dwarfs 
may have formed after the cluster
was in place.  There is now evidence for this, the implications of
which can explain several galaxy cluster phenomenon.  On the other
hand, there is also evidence that some dwarf ellipticals/spheroidals
in clusters are dominated by old stellar populations.  In this paper we 
review the current observations of dwarfs in clusters and attempt to 
interpret these systems in terms of known properties of Local Group
dwarf ellipticals/spheroidals, and in the milieu of
theoretical ideas concerning low
mass galaxy formation in a cosmological context.

\section{Cluster Dwarf Galaxy Properties}

Dwarfs have been studied in detail in nearby
rich clusters, such as Virgo, Fornax and Coma, and Perseus.  Most
of the data we discuss therefore comes from these sources.  In particular
a significant fraction of the data presented in this article comes from the
papers by
Conselice, Gallagher \& Wyse (2001,2002,2003) and Conselice et al. (2003b).
We list below our current understanding of dwarf properties in 
terms of various observational quantities.

\vspace{0.2cm}
\noindent {\bf Number Densities}:  The
luminosity function (LF) in all galaxy environments are dominated
by dwarf galaxies. The over-density of dwarfs in clusters
is about 5-10 times that in groups, that is the dwarf to giant ratio
is nearly a factor of ten higher.  Another way to quantify this is through
the faint end slope of the LF, $\alpha$.  The value
of $\alpha$ in rich clusters, such as Virgo is typically
around $\alpha = -1.2$ to $ -1.5$ (Sandage et al. 1985; 
Trentham et al. 2002), with some results suggesting even steeper LFs with
$\alpha = -1.6$ (Sabatini et al. 2003).   This is steeper than field
values, such as in the Local Group ($\alpha = -1.1$; 
van den Bergh 1992), yet flatter than the value predicted by CDM for all
environments ($\alpha = -2$).  Environment
clearly affects the way these systems are produced, which is generally
not a prediction in CDM models (cf. Tully et al. 2002).  This also
suggests that clusters of galaxies {\it cannot} simply form from the
mergers of lower density groups of galaxies.

\vspace{0.2cm}
\noindent {\bf Spatial Positions}: While Local Group dwarf galaxies, 
particularly
dwarf ellipticals, are strongly clustered around giant galaxies
in the Local Group, the opposite is found for 
low-mass galaxies in clusters, where most are neither clustered around, nor 
distributed globally similar to,  giant elliptical galaxies 
(Conselice et al. 2001).  Both dwarf ellipticals and irregulars also
have a broader distribution within clusters, that is they are not clustered
towards the center, but are spread throughout 
(e.g., Conselice et al. 2001).  This is also the pattern seen for
spirals and irregulars in clusters.

\begin{figure}
\hspace{-0.7cm}
\includegraphics[height=3.5in,width=6in,angle=0]{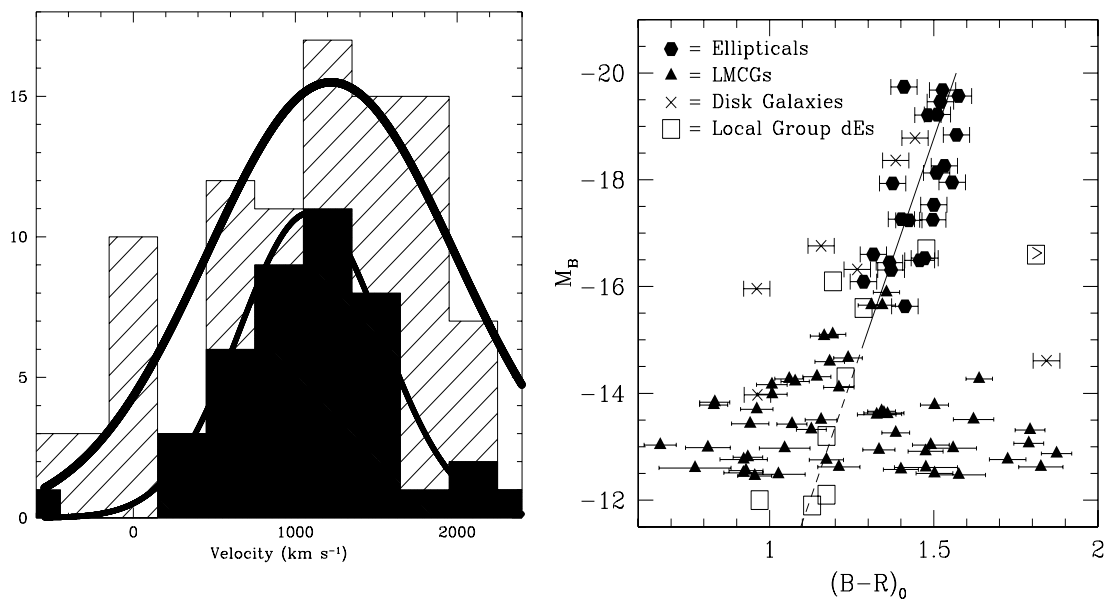}
  \caption{(a) Velocity histograms for giant ellipticals (solid) and dwarf 
ellipticals (shaded) in the Virgo cluster (Conselice et al. 2001). (b) 
Color magnitude diagram for galaxies in the Perseus cluster, 
demonstrating the large color scatter for systems with M$_{\rm B} > -15$.
The solid boxes are where Local Group dEs/dSphs would fit on this plot. Dwarf
ellipticals are labeled as low mass cluster galaxies (LMCGs).} \label{fig:wave}
\end{figure}

\vspace{0.2cm}
\noindent{\bf Radial Velocities}: The radial velocities of low-mass cluster 
galaxies, including S0s, spirals, dwarf irregulars and dwarf
ellipticals have a wider distribution than the ellipticals (see
Figure~1a).  For example, Virgo cluster elliptical
galaxies have a narrow Gaussian velocity distribution, with $\sigma = 462$ 
km s$^{-1}$, concentrated at the mean 
radial velocity of the cluster.  The other populations, including
the over 100 classified dwarf ellipticals in Virgo with radial
velocities, have much broader, and non-Gaussian, velocity distributions 
($\sigma \sim 700$ km s$^{-1}$), all with velocity dispersion ratios with 
the ellipticals consistent with their being accreted 
(e.g., Conselice et al. 2001).  There is also significant sub-structure within
these velocity distributions, unlike the case for the giant
ellipticals.

\vspace{0.2cm}
\noindent{\bf Stellar Populations}:   Currently,
we know with some certainty that dwarf galaxies have either 
young/metal rich or old/metal poor stellar populations 
(e.g.,  Poggianti et al. 2001).  This is further seen
in complete color-magnitude diagrams in nearby rich clusters,
such as Perseus,  down to M$_{\rm B} = -12$ (Conselice et al. 2002, 
2003a).  Fainter dwarfs also tend to be even more heterogeneous, with a large
scatter from the color-magnitude relationship (CMR) (Conselice et al.
2003a; Rakos et al. 2001) (Figure~1b).  This trend is found in
several nearby clusters, including Fornax, Coma and Perseus, and can be 
explained by different dwarfs having mixtures of ages and 
metallicities (e.g., Poggianti et al. 2001; Rakos et al. 2001; 
Conselice et al. 2003a).  Stromgren
and broad-band photometry reveal that the redder dwarfs are metal enriched
systems (Figure~2).   However, old stellar populations do exist in
dwarf galaxies (Poggianti et al. 2001) and in the globular clusters
surrounding dwarfs (e.g., Conselice 2005).

\vspace{0.2cm}
\noindent {\bf Internal Kinematics}:    One key observational test for
the origin of dwarf ellipticals is whether they rotate or not. Models
show that a dwarf
elliptical which has been transformed from a spiral should reveal some
rotation.  Using 8-10 meter class telescopes, the evidence is ambiguous
with some dEs showing rotation (Pedraz et al. 2002), while others
clearly do not (Geha et al. 2003).  There is no obvious difference between
these two populations in terms of morphology or chemical abundances 
(Geha et al. 2003), although these samples are still very small.  The
central velocity dispersions of these systems is also quite low, indicating
that dark matter does not dominate, at least in their centers.

Furthermore, in addition to predicting low-metallicity systems, Dekel \& Silk (1986) show, based on their model assumptions, how the internal velocities of 
low-mass galaxies should correlate with M$_{\rm tot}$/L ratios and 
luminosities, L.  The relationship they predict is M$_{\rm tot}$/L $\sim$ 
L$^{-0.37}$, and $\sigma \sim$ L$^{0.19}$.  These predictions can be 
tested utilizing Virgo 
Cluster dE internal velocity measurements made by Pedraz et al. (2002) and 
Geha et al. (2003).  After fitting a power law to the relationship between 
$\sigma$ and L,  $\sigma \sim$ L$^{0.31\pm0.05}$. The fitted exponent on L is 
2$\sigma$  away from the relationship predicted by Dekel \& Silk.  This is 
another indication that cluster dwarf ellipticals are possibly formed through
multiple methods.

\vspace{0.2cm}
\noindent {\bf Gas Content}:  Surprisingly, some dEs in clusters contain
evidence for gas in various phases, both cold HI (Conselice et al. 
2003b; Buyle et al. 2005) and warm gas in the form of H$\alpha$ (Michielsen
et al. 2004).  These studies
have only been performed in the nearest clusters, Virgo and Fornax due 
to the difficulty of making these observations.  Although 15\% of
dEs have evidence for HI, all of these systems are located in the outer
parts of Virgo (Conselice et al. 2003b) and Fornax (Buyle et al. 2005).  
Figure~3 shows the location of HI emitters in the Virgo cluster and the 
fraction of dEs with HI emission as a function of radius from the center of
the Virgo cluster.

\begin{figure}
\hspace{-0.2cm}
\includegraphics[height=6.5in,width=2.5in,angle=-90]{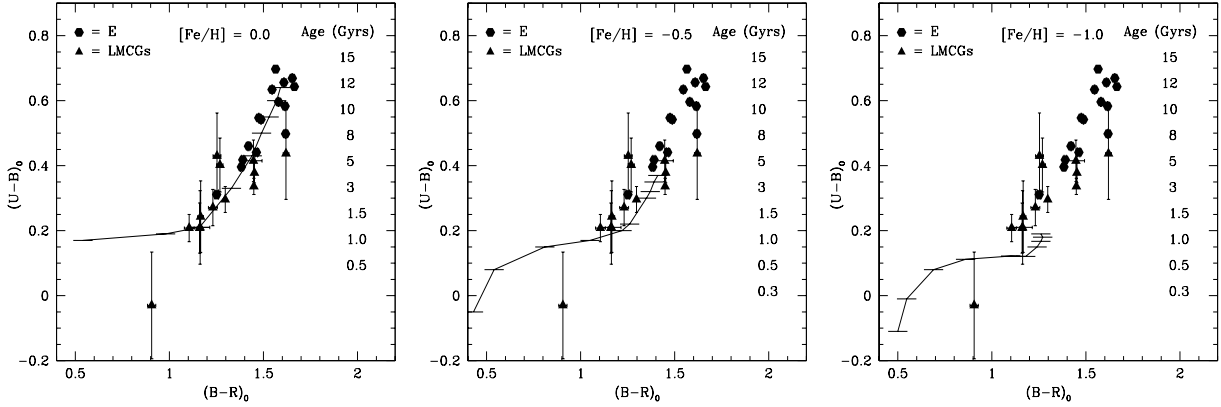}
  \caption{Three stellar synthesis modeled age tracks on a UBR color-color 
diagram 
at constant metallicities of solar, [Fe/H] = -0.5, and [Fe/H] = -1. The age
range is 0.3 Gyrs to 15 Gyrs for the [Fe/H] = -0.5 and -1
models and 0.5 Gyrs to 15 Gyrs for the solar metallicity models. Dwarf
ellipticals are labeled as low mass cluster galaxies (LMCGs).  } \label{fig:wave}
\end{figure}

\begin{figure}
\hspace{-0.7cm}
\includegraphics[height=3in,width=6in,angle=0]{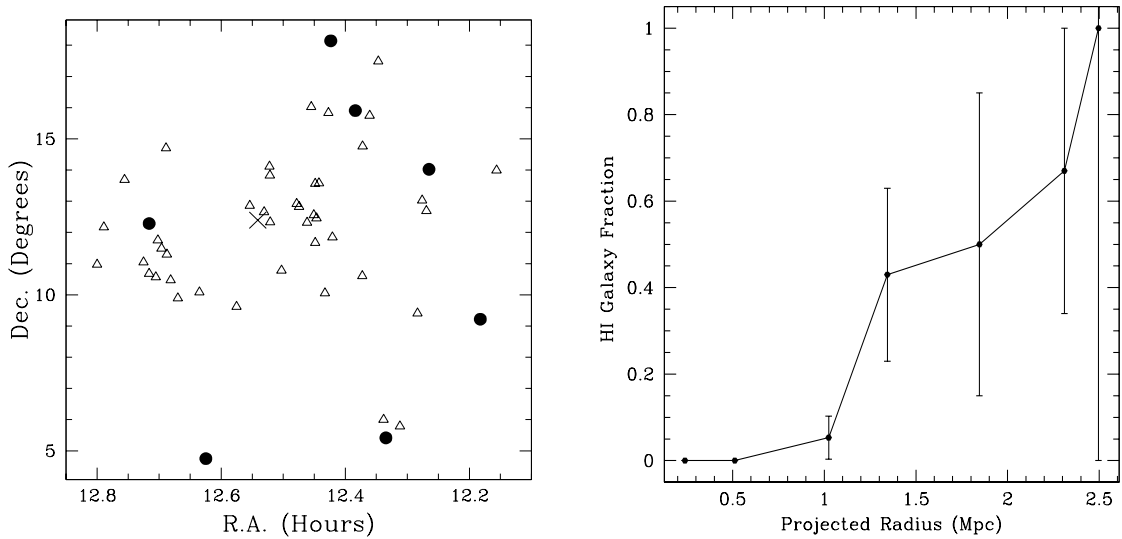}
  \caption{ (a) Distribution on the sky of dE galaxies observed at 21 cm in the Virgo Cluster. The open triangles represent non-detections, while the filled circles show the locations of the dE galaxies with detected H I. The cross toward the center shows the location of the giant elliptical M87.  (b) Fraction of Virgo classified dwarf elliptical galaxies detected in H I as a function of projected distance from the center of the cluster.  } \label{fig:wave}
\end{figure}

\section{Cluster Dwarf Galaxy Origins}

Any successful theory of dwarf galaxies, particularly for explaining how
dwarf ellipticals form, must account for the following properties: 
over-density in
relation to massive galaxies in rich environments, mix of stellar populations,
ability to survive in dense environments, diffuse spatial and velocity
structure, and mixtures of rotation.  This theory must also explain why
fainter dwarfs are more heterogeneous than brighter ones.

In the simple collapse + 
feedback scenarios (Dekel \& Silk 1986), dwarfs are formed when gas collapses 
and forms stars. These stars produce winds that expel gas from these systems,
halting any future star formation.
In this formation scenario dwarfs formed before the cluster
ellipticals, or at least formed within groups that later merged to form 
clusters. Fainter dwarfs 
however, cannot all be born in groups which were later accreted into 
clusters along with the massive galaxies, due to the high dwarf to giant 
galaxy ratio found in clusters (Conselice et al. 2001, 2003a).

One idea to explain this, as suggested by Tully et al. (2002), is that 
the dark matter
halos of dwarfs are `squelched' in lower density environments due to a large
ultraviolet background after the universe was reionized.  This explains the
differences in $\alpha$ between different environments, but does not explain
how within the same environment there is a great diversity in the dwarf
population. It also does not easily explain why many dwarfs appear to have
recently been accreted into clusters.   

The above evidence suggests that simple low-mass galaxy formation scenarios
can be safely ruled out for some dwarfs.  
The first clue that dwarfs are not produced through standard methodologies 
came from studies that showed dwarfs are generally found in abundance in 
dense areas such as nearby clusters.   In fact, in standard CDM scenarios
dwarfs should be more common in lower density environments, but we see
the direct opposite (Trentham et al. 2002).  Dwarfs within dense
environments also have a broader distribution, both spatially and in terms of
their radial velocities, than giant ellipticals, similar to 
the pattern of infalling spirals.  This is an indication that both are 
recent additions to clusters (Conselice
 et al. 2001).  Internal dynamic evidence also suggests that at least some
dwarfs are rotating (e.g., Geha et al. 2003), a feature not seen in Local
Group dEs.  Finally, the stellar populations of some faint M$_{\rm B} > -15$ 
dwarf 
galaxies appear to be metal rich (near solar), implying that the dwarf
population itself is inhomogeneous and may have multiple origins.  We discuss
several scenarios which have been proposed to explain these observations and
argue that one method which can reproduce these trends involves
the formation of dwarfs from already existing galaxies through a tidal effect.

One alternative idea is that some modern dwarfs
formed after the cluster itself was in place by collapsing out
of enriched intracluster gas.  Another is that the intracluster medium (ICM)
is able to retain enriched gas that in the Dekel and Silk
(1986) paradigm would be ejected by feedback, but remains due to the 
confinement pressure of the ICM (Babul \& Rees 1992).  This scenario would 
explain the higher metallicities of some of the fainter dwarfs.

An alternative scenario, now gaining in popularity, is that dwarfs form in 
clusters through
a tidal origin.   Two main possibilities for this are tidal dwarfs (Duc
\& Mirabel 1994), and as the remnants of 
stripped disks or dwarf irregulars (Conselice et al. 2003a).
The velocity and spatial distributions of dwarfs suggests that they were likely
accreted into clusters during the last few Gyrs 
(Conselice et al. 2001).  This, combined
with the high metallicities of these cluster dwarfs, and the fact that their
stellar populations are fundamentally different than field dwarfs (e.g.,
Conselice et al. 2003a; Figure~1b; Figure~2) suggests that the cluster environment has
morphologically transformed, or stripped, accreted galaxy material into dwarfs.

There is evidence for this process currently occurring in nearby clusters 
(e.g., Conselice \& Gallagher 1999).  
If dEs form from infalling spirals then there should also
be systems now being transformed which appear morphologically as dEs, but
retain some of the gas left over from their precursor.  These systems
would only exist in the outer parts of clusters, as any that dwarfs
venturing towards the
core will be rapidly stripped of material.  For example, very deep Arecibo 
observations
of Virgo dEs reveal that $\sim 15$\% of a sample of 56 have
HI detections, all of which are located outside the core of
the cluster (Conselice et al. 2003b).  Other detailed morphological 
investigations of nearby cluster
dwarf ellipticals show that they contain a wide diversity of structures,
some with tidal features, and others with apparent spiral structures (e.g., 
Jerjen et al. 2000; De Rijcke et al. 2003; Barazza et al. 2003; Graham et al. 2003).

Despite the above, there is considerable evidence that some dEs in
clusters are indeed an old population, some with metal poor populations
(e.g., Lotz et al. 2004). A single cluster dwarf scenerio is unlikely
to explain the great diversity we see, thus multiple formation methods are
likely necessary.

\section{Implications of Late Dwarf Formation in Clusters}

Environment likely plays a role in all aspects of galaxy formation and
evolution. If indeed any dwarfs form after their host cluster due to
a tidal process, there are several cluster features these processes can 
possibly explain.  The first is that the LF of clusters will change
after these lower mass galaxies form.   The LF of
galaxies in the Perseus cluster becomes flatter, with a similar
faint end slope ($\alpha$) as the field, after removing these red galaxies 
(Figure~4).   Figure~4 shows how the luminosity function changes slope
from $\alpha = -1.4$ to $\alpha = -1.2$ due to galaxies in the Perseus
cluster which are redder than 2$\sigma$ from the color-magnitude relation,
defined by the giant cluster galaxies (Conselice 2002).

\begin{figure}
\hspace{-0.7cm}
\includegraphics[height=2.5in,width=6in,angle=0]{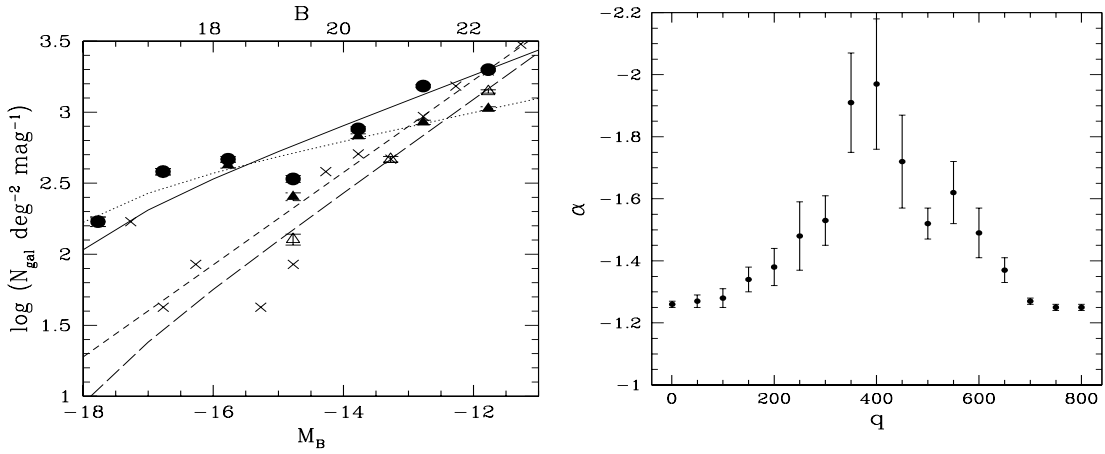}
  \caption{(a) The Perseus cluster luminosity function plotted in
various ways down to $M_{B} = -11$.  The solid round points, 
and fitted solid line, is the total luminosity function of the central 
region 
of Perseus.  The luminosity function for dEs redder and bluer than the CMR
prediction are plotted as open and solid triangles and long dashed and
dotted lines.  The crosses mark the density of background galaxies. (b)
Modeled cluster luminosity function
slope, $\alpha$, as a function of the number of high-speed maximum 
interactions 
(q $\sim$ time) cluster galaxies undergo during evolution in a Perseus like 
cluster (see Conselice 2002).  } \label{fig:wave}
\end{figure}

There are several implications for this process, beyond possibly steepening
the luminosity function.  One of these is the origin of intracluser
light, which can make up 50\% of the light coming from dense clusters
(e.g., Adami et al. 2005).  If cluster dEs originate from tidally disturbed 
galaxies then the amount of light liberated is enough to account for all 
of this intracluster
light as arising from the debris from tidally disturbed galaxies
(Conselice et al. 2003a).  For example, the total luminosity of 
intracluster light in the Virgo Cluster within 2$^{\circ}\,$ of M87 is 
2 $\times 10^{11}$ L$_{\odot}$. If this material originates from tidally 
striped  objects whose remnants are dwarfs, we can compute how much material 
on average each dE must have lost.  There are 170 dEs, 148 dE,Ns, and 14 S0 
galaxies within  this radius in Virgo. On average, if all the dE and dE,N 
galaxies are remnants of stripped galaxies then 
$0.6 \times 10^{9}$ L$_{\odot}$ of light was lost by each object. 
If we consider only dE galaxies as remnants of this process then $1.1 \times 
10^{9}$ L$_{\odot}$ was lost by each dE.  Therefore, if we assume
that dEs were  once as massive as a typical disk galaxy, then the amount of
material these systems lost through tidal effects is enough to account for
the intracluster light in rich clusters of galaxies.

There are still however many questions that need to be answered by
observing dwarfs in clusters in more detail.  There has never been a
spectroscopic survey of dwarfs fainter than M$_{\rm B} = -15$ in clusters
of galaxies, for example.  We are also just learning about the internal
properties of dwarfs through high resolution imaging and spectroscopy.  Making
progress on these topics will require surveys of dwarfs in all
environments, and connecting these observations
with galaxy evolution in higher redshift clusters.

\acknowledgments{I acknowledge my collaborators for their participation in the projects described here. I also
thank the organizers of this symposium for creating an interesting meeting, and
for their patience in receiving this review.

\end{document}